# A Cosmological Modification of Relativistic Energy from Mach-Hamilton Consistency

*Scott Funkhouser*
*Occidental College, Los Angeles, CA 90041*
23March2004

**ABSTRACT**
If Mach's Principle explains the Newtonian inertial reaction to acceleration then the role of the "fixed stars" should also be manifest through Hamilton's formulation of mechanics. This consistency may be achieved if the expression for relativistic energy contains a cosmological coefficient that is (currently) equal to one. The presence of the required cosmological term exactly identifies the rest energy of a body as its gravitational potential energy due to the mass of the universe.

## 1. Mach's Principle and Sciama's Force Law

According to Mach's Principle, Newton's Second Law of Motion is the result of an instantaneous reaction between any accelerated mass and the mass of the universe. The applied force required to counter the inertial resistance of a body to acceleration would be explained if the gravitational interaction between any two masses contains a new component proportional to the relative acceleration between the masses. The additional force between any two masses $m_1$ and $m_2$ that would account for inertia was detailed by Sciama [1] (among others) and is given by

$$\vec{F}_a = -\frac{Gm_1 m_2}{r_{12}} \frac{\vec{a}_{12}}{c^2} \tag{1}$$

where $a_{12}$ is the relative acceleration of the two masses, $r_{12}$ is their separation, $G$ is Newton's gravitational constant and $c$ is the speed of light. This force becomes non-negligible when evaluated with respect to a very large mass. When evaluated between a local mass $m$ and the mass of the Universe $M_U$, the acceleration-dependent force becomes

$$\vec{F}_a = -A \frac{GM_U}{R_U c^2} m\vec{a} \tag{2}$$

where $A$ is an integration factor of order unity, $R_U$ is the radius of curvature of the Universe and $\vec{a}$ is the acceleration of the local mass with respect to the fixed stars. This force $\vec{F}_a$ could represent the Newtonian force $m\vec{a}$ if $AGM_U R_U^{-1} c^{-2} \equiv \varepsilon_U$ is equal to one. This relationship is indeed satisfied within order of magnitude if $R_U \sim 10^{26}$m and $M_U \sim 10^{53}$kg. (Note that the Newtonian force is understood to *counteract* the inertial resistance to acceleration, thus the negative sign is omitted in the Newtonian formulation.)

This approach suggests that the Newtonian equation of motion could be stated more fully as

$$\vec{F} = \varepsilon_U m\vec{a} \tag{3}$$

where the cosmological coefficient $\varepsilon_U$ is currently equal to one. Whether or not this coefficient is era-dependent is one important question among many which must be addressed eventually by any successful formulation of Mach's Principle. This present work addresses a separate issue associated with the Sciama formulation of Mach's Principle, namely a consequence of consistency between Mach's Principle and the Hamiltonian treatment of classical dynamics.

## 2. Consistency with Hamilton's Principle

The inertial resistance to acceleration produced by Sciama's formulation of Mach's principle ought to be identical to the result obtained from Hamilton's variational principle for the dynamics of a body

$$\delta \int_{t_1}^{t_2} L\{q_i, \dot{q}_i; t\} dt = 0 \tag{4}$$

where $L$ is the Lagrangian of the body expressed in terms of the generalized coordinates $q_i$. The extremum condition of Eq. (4) leads to the Lagrange-Euler equations of motion

$$\frac{\partial L}{\partial q_i} - \frac{d}{dt}\frac{\partial L}{\partial \dot{q}_i} = 0. \tag{5}$$

In the most basic non-relativistic application the Lagrangian $L$ for a mass $m$ moving along one Cartesian axis $\hat{x}$ is

$$L = \frac{m}{2}\dot{x}^2 - U(x) \tag{6}$$

where $U(x)$ is the potential energy of the body. The Lagrange-Euler equation applied to this case leads to

$$-\frac{\partial U}{\partial x} - \frac{d}{dt}(m\dot{x}) = 0. \tag{7}$$

Identifying the first term on the left as the force $F$ acting on the body, Eq. (7) immediately reproduces Newton's Law $F = m\ddot{x}$ [2].

However, if Mach's Principle is a correct representation of nature then the result of the Lagrange-Euler equation(s) should contain implicitly some accounting of the action of the mass $M_U$ of the Universe. This may be achieved most naturally if the kinetic term in the Lagrangian is multiplied by the "invisible" coefficient $\varepsilon_U$. If the non-relativistic kinetic energy $T$ of a mass $m$ with velocity $\dot{x}$ should be given by

$$T = \varepsilon_U \frac{m}{2}\dot{x}^2 \tag{8}$$

then the Lagrange-Euler method would produce a law of motion identical to that obtained from the Sciama force law and Mach's Principle in Eq. (3).

## 3. The Relativistic Rest Energy and the Cosmos

Consistency with the special theory of relativity requires that the energy $E$ of a mass $m$ traveling with velocity $\beta c$ also contain the cosmological coefficient $\varepsilon_U$

$$E = \varepsilon_U mc^2 (1-\beta^2)^{-1/2}. \tag{9}$$

In the limit $\beta \ll 1$ this gives the non-relativistic expression required by Eq. (8) (within an additive constant rest energy). The presence of such a coefficient in the expression for energy may bear implications for interpreting the rest mass-energy of any body. The rest energy would still contain the cosmological term and would be given explicitly as

$$\varepsilon_U mc^2 = A\frac{GM_U m}{R_U} \tag{10}$$

which is to say that the relativistic rest energy of any mass is equivalent to (the absolute value of) its gravitational potential energy in the Universe. This correspondence between the cosmological gravitational energy and the rest energy has been noted before, but previous arguments have been based on the substitution $c^2 \sim GM_U/R_U$ into the expression $E=mc^2$. Equation (10) constitutes an exact relationship with the speed of light eliminated algebraically from the expression for rest energy regardless of any presumed physical relationship between $c$ and the mass distribution of the cosmos.